\documentclass[prc,twocolumn,nofootinbib]{revtex4-2}

\usepackage{color}
\usepackage{graphicx}
\usepackage{dcolumn}
\usepackage{bm}
\usepackage{braket}
\usepackage{amsmath}
\usepackage{here}
\usepackage{color}
\usepackage{enumerate}
\usepackage{amssymb}
\usepackage{bbm}
\usepackage{dsfont}
\def\uf{$^{236}$U}

\def\be{\begin{equation}} 
\def\ee{\end{equation}} 
\def\unit{\mathds{1}}

\usepackage{xcolor}

\usepackage[normalem]{ulem}  

\renewcommand\sout{\bgroup\color[rgb]{1,0.75,0.8} \ULdepth=-.5ex \ULset}

\renewcommand{\vec}[1]{\mbox{\boldmath $#1$}}

\begin{document}

\title{Schematic model for induced fission in a configuration-interaction approach}

\author{K. Uzawa and K. Hagino}

\affiliation{%
 Department of Physics, Kyoto University, Kyoto 606-8502, Japan
}%

\begin{abstract}

We model fission at barrier-top energies in a simplified model space that permits
comparison of different components of the residual nucleon-nucleon interaction.
The model space is built on particle-hole excitations of reference configurations.
These are Slater determinants of uniformly spaced orbitals characterized only by 
their quantum numbers and orbital energies.  The residual interaction in the
Hamiltonian includes the diabatic interaction connecting similar orbitals at
different deformations,  the pairing interaction between like nucleons, and 
a schematic off-diagonal neutron-proton interaction.     
We find that the fission reaction probability is 
sensitive to the off-diagonal neutron-proton interaction much more than to the pairing 
and the diabatic interactions. In particular, the transmission coefficients become insensitive to the strength 
of the pairing interaction when the neutron-proton interaction is large. 
We also find that the branching ratio is  insensitive to  
the final-state scission dynamics, as is assumed in the well-known Bohr-Wheeler theory. 
\end{abstract}

\maketitle

\section{INTRODUCTION}

Nuclear fission was discovered about 80 years ago \cite{hahn1939,va73}. 
Many phenomenological models have been proposed since then and 
have successfully
explained the observed behaviors.  A well-known model is of Bohr and Wheeler
\cite{bohr1939}, in which a statistical treatment is implemented under the
transition-state hypothesis.  In addition to this model, the statistical
models based on the Hauser-Feshbach theory\cite{hauser1952} as well as
dynamical models based on a transport
theory\cite{randrup2011,aritomo2014,ishizuka2017} have also played an
important role \cite{schmidt2018}.  In contrast, a microscopic understanding
of induced fission has still been far from complete.  This has been regarded
as one of the most challenging subjects in many-fermion quantum dynamics,
and in fact in a recent review for future directions of fission
theory\cite{bender2020}, the authors omitted this topic “because there has
been virtually no coherent microscopic theory addressing this question up to
now.”

In this paper, we apply the configuration-interaction (CI) approach
\cite{CI2022,BH2023} to a schematic model in order to discuss the role of
various types of nucleon-nucleon interaction.  In this approach,
many-particle-many-hole configurations at different nuclear deformations are
coupled by residual interactions.  Those many-body configurations are
constructed in a constrained mean-field potential at each deformation. 
The configuration space includes particle-hole excitations of the
reference configurations and thus greatly extends the space accessed by the
collective coordinates defined in the usual generator coordinate method
(GCM)\cite{ring}. See Ref.  \cite{donau1989} for a similar approach.

In a recent publication \cite{BH2023}, the CI approach was applied to
semi-realistic calculations based on the Skyrme energy functional.  However,
for simplicity, several simplifications were introduced.  In particular, the
model space was restricted to neutron excitations only with seniority zero. 
As a consequence, only two types of interaction were needed, namely the
pairing and the diabatic interactions.  In nuclear structure the
off-diagonal neutron-proton interaction is important as well, but its role in
low-energy nuclear fission has not yet been clarified.  

In this paper, we apply the CI approach to a schematic model with 
uniformly spaced single-particle orbitals. A preliminary
version of the work can be found in Ref.  \cite{CI2021A}; some of the
supplementary material of that work is included in Appendix A of this paper. 
While the model presented here is still far from realistic, our schematic
treatment of the configuration space and the details of the Hamiltonian may
be useful for focusing  attention on aspect of those ingredients in more
quantitative theory.  This is especially needed in light of the huge CI
spaces required to describe the large changes of deformation that occur in
fission.

The paper is organized as follows.
Sec. II presents the theoretical framework and the model Hamiltonian based on uniformly
spaced orbital energies.  
In Sec. III we apply the  model to transmission across a barrier.  There are three kinds
of residual interaction that can mediate the transmission dynamics, and we examine their
relative importance. The interaction types are diabatic, pairing, and the fully off-diagonal
nucleon-nucleon interaction.  Taking the model as a schematic treatment of fission, 
we examine in Sec. IV the  branching ratio between fission and the capture. It
is shown that one of the tenets of the 
Bohr-Wheeler theory (insensitivity to fission partial widths) can be achieved 
with the model Hamiltonian. 
We then summarize the paper in Sec. V. 

\section{CI approach to induced fission}

\subsection{Transmission coefficient}

In the present approach, the reference configurations are defined at
discrete points along the fission path. Many-particle-many-hole excited
states are then generated from those reference configurations to form
subspaces in the configuration space which we call $Q$-blocks.
In general, the states in different
$Q$-blocks are not orthogonal to each other, and one needs to consider the
norm 
matrix $N$ with matrix elements $\langle q j | q' j' \rangle $ where
$q$ and $q'$ label the $Q$-blocks and $j,j'$ are labels for the
configurations within a $Q$-block.  Similarly, the
elements of Hamiltonian  matrix $H$ are $\langle q j | H | q'
j'\rangle$.
Note that in the usual GCM one takes only the local ground state at each $q$.  In
contrast, induced fission is a decay process of an excited nucleus, and it
is essential to include excited configurations.  

The $S$-matrix reaction theory also
requires matrices $\Gamma_c$ for decay widths to each channel $c$.  Note
that these matrices have the same dimension as the configuration space 
and may have off-diagonal matrix elements.
  The present model includes neutron entrance channels, multiple
$\gamma$-ray capture channels, and multiple fission channels; the
corresponding matrices
are $\Gamma_n$,
$\Gamma_{\rm cap}$, and $\Gamma_{\rm fis}$
respectively.  Specific forms of those matrices are given in Sec.  IID
below.

Based on the Datta formula from $S$-matrix reaction theory \cite{datta,al20,al21}, 
we evaluate the transmission coefficient from the incoming channel 
$a$ to a decay channel $b$ at energy $E$ as
\begin{equation}
T_{a,b}(E)=\sum_{i \in a,j \in b}|S_{i,j}(E)|^2=\mathrm{Tr}[\Gamma_aG(E)\Gamma_bG(E)^\dagger],
\label{Tab}
\end{equation}
where 
\begin{equation}
G(E)=(H-i(\Gamma_n+\Gamma_{\rm cap}+\Gamma_{\rm fis})/2-NE)^{-1}
\label{green}
\end{equation}
is the Green's function with the total width $\Gamma=\Gamma_n+\Gamma_{\rm cap}+\Gamma_{\rm
fis}$.  

In a low-energy induced fission, the channel $a$ corresponds to the
incident channel and thus $\Gamma_a=\Gamma_n$, while the exit channel $b$ is
either the capture channel or the fission channel.

In GCM calculations for nuclear spectroscopy, it is well known that the
non-orthogonality of a basis set often leads to a numerical instability
\cite{ring,martinez-larraz2022}.  One can largely avoid this problem in
reaction calculations, as a rather coarse mesh along the fission path
 provides an acceptable accuracy for estimating the transmission coefficients \cite{CI2022}.

\subsection{Model Hamiltonian}

The Hamiltonian for each $Q$-block is constructed as
\begin{equation}
H_q=V(q)+H_{\rm sp}+H_{\rm pair}+H_{\rm ran}, 
\label{hamiltonian}
\end{equation}
where $V(q)$ is the energy of the local ground state at $q$.
Ideally, it is 
calculated by constrained Hartree-Fock or density functional
theory (DFT).  $H_{\rm sp}$, $H_{\rm pair}$, and $H_{\rm ran}$ are the
single-particle Hamiltonian, the pairing interaction, and the random
neutron-proton interaction, respectively.

The configuration space is built in the usual way, defining configurations
as Slater determinants of nucleon orbitals.  
In this
paper, we employ a model having  a uniform spectrum of orbital energies 
with a spacing $d$ for both 
protons and neutrons.  The ladder of orbital states
extends infinitely in both directions above and below the Fermi 
surface, but is restricted by later trucation of the CI space.
The operator for the particle-hole excitation energy $E_{ph}$ is given by
\begin{equation}
H_{sp}=d\sum_{\alpha:n_a>0} n_{\alpha}a^\dagger_{\alpha}a_{\alpha}+
d\sum_{\alpha:n_a<0} n_{\alpha}a_{\alpha}a^\dagger_{\alpha}.
\end{equation}
The label $\alpha$ includes $q$ and an index $\nu$ of orbitals associated
with the reference configuration.  The interaction matrix elements
also require access to the conserved quantum numbers of orbitals.  In
general, these include parity $\pi$, isospin $t_z$, and angular momentum
$K$ about the symmetry axis if there is one.  
To keep the model as transparent as possible, we ignore parity and
assign $K$ to the restricted range  $\pm1/2$.   

The orbital excitation energies of many-particle configurations
are integral multiples of $d$, given by $E_{\rm{ex}} = k d$.  As a
function of $k$, the multiplicity of configurations having
$K_{\rm tot}\equiv\sum K = 0$ and $\sum t = 0$ is $N_k =
(1, 4, 16, 48, 133, 332, 784, \cdots)$ for $k = (0, 1, 2, 3, 4,
5, 6, \cdots)$.  The spectrum up to $k=6$ is shown in Fig.  1. 
The orange curve shows a smoothed level density fitted to the
leading order dependence on energy as derived from statistical
theory.  This will provide a way to fit the parameter $d$ to
experimental level densities: the single-particle level spacing
$d$ sets the energy scale in the model, and other energy
parameters will be expressed in units of $d$.  Even though we
will not specify the value of $d$ in this paper, $d$ is
estimated to be around 0.5 MeV for nuclei in the actinide region
\cite{CI2021A} (see Appendix A-1).

\begin{figure}
\includegraphics[width=8.6cm]{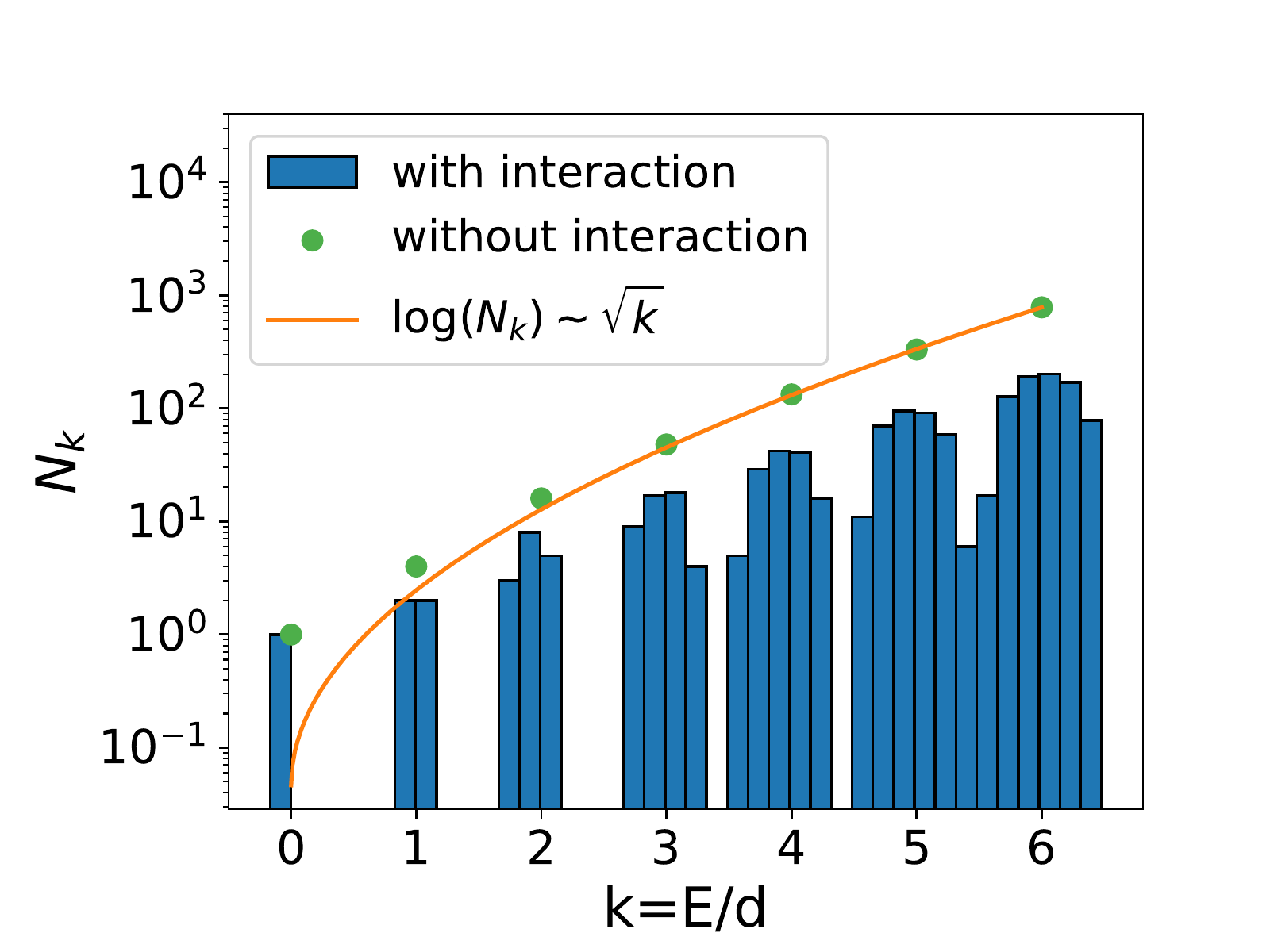}
\caption{Spectrum of many-body configurations in the uniform spacing model. 
$N_k$ denotes the number of $K_{\rm tot}\equiv\sum K = 0$ configurations at the excitation energy $E=kd$. 
The green circles show the non-interacting spectrum, while the orange curve
shows its fit to the functional form of $N_k=\exp(a\sqrt{k}+b)$ with
$a=3.97$ and $b=-3.06$.  The blue filled histograms show the interacting
spectrum, obtained by diagonalizing the Hamiltonian $H_q$ with $G_{\rm pair}=0$ and
$v_{np}=0.03d$.
}
\end{figure}

For residual interactions, both particle-particle (pp) and particle-hole (ph) 
interactions appear.
For the pp residual interaction, we employ a monopole pairing interaction
between identical nucleons, 
\begin{equation}
H_{\rm pair}=-G_{\rm pair}\sum_{\nu\neq\nu'}
a^\dagger_{\nu'} a^\dagger_{\bar{\nu}'} 
a_{\bar{\nu}}a_\nu. 
\end{equation}
Here $a^\dagger_{\nu}$ is the creation operator of the orbital
$\nu$, and 
${\bar{\nu}}$ denotes the time-reversal orbital of $\nu$. 
The strength of the pairing interaction $G_{\rm pair}$ is around $0.1$ MeV in the
actinide region \cite{superfluidity}, corresponding to $G_{\rm pair}\approx 0.2 d$ in
the energy units in the present model.  In this paper we take $G_{\rm pair}=0.3d$ as
the baseline value, to be varied to study how the observables depend on the
interaction types.

When the monopole pairing interaction is used in the uniform spacing model, 
the Hamiltonian matrix tends to be singular 
due to the high degeneracy of the spectrum. In actual numerical calculations, 
an unphysical divergent behavior 
may easily appear in the transmission coefficients even if the matrix itself is invertible.
To avoid this problem, we add a small random number to the diagonal
part of the Hamiltonian kernel as \cite{CI2020} 
\[kd\rightarrow kd+0.1\,r\,kd,\]
where $r$ is a random number of unit variance taken from a Gaussian ensemble.

For the ph-type residual interaction, we employ a random interaction in the form of 
\begin{equation}
H_{\rm ran}=-v_{np}\sum^{'} r a^\dagger_{\alpha1} a^\dagger_{\alpha2}  a_{\alpha4}a_{\alpha3},
\end{equation}
where the parameter $v_{np}$ is the strength of the interaction and $r$ is a random
number,
as before. 
The sum $\alpha$ is restricted to the combinations satisfying
$K_1+K_2=K_3+K_4$.  An early study has suggested that a neutron-proton
interaction is dominant in a diffusion process compared to the one between
identical particles\cite{bush1992}.  We therefore assume that the
interaction $H_{\rm ran}$ acts only on neutron-proton pairs.

Following Appendix \ref{A3}, we take $v_{np}=0.03 d$ as a base value in the
following calculations.  The assumption that the neutron-proton interaction
is Gaussian distributed is certainly not justified for the low-energy states
in a $Q$-block where collective excitations can be built up.  However, high
in the spectrum 
the mixing approaches the random matrix limit.  Note that the pairing
interaction acts coherently while the random interaction acts incoherently. 
Our interest is to clarify the role of these two different types of
interaction in the transmission process.

Because of the random component  in the Hamiltonian, 
one needs to take an ensemble average to obtain physical quantities. 
In the following calculations, we take many samples so that the standard 
deviation becomes smaller than 1\%. 

\subsection{Off-diagonal couplings}

The interaction between different $Q$-blocks is responsible for a shape change and is  
thus crucial to the modeling.
It is clear that the interaction is somewhat suppressed due to the imperfect 
overlap of orbitals built on different mean-field reference states. 
The size of the suppression is determined by the overlap kernel,
$N(qj,q'j')$, 
which is given by a determinant of orbital overlaps.
For simplicity, we assume that the change of the single-particle orbitals 
between nearby reference configurations is small. 
With this assumption, the configurations with the same index $j$  in neighboring $Q$-blocks 
are diabatically connected with the overlap matrix elements approximated by
\begin{equation}    
\langle q\, j |N| q'\, j'\rangle = N(q,q')\delta_{j,j'},
\end{equation}
where $N(q,q')$ is the overlap between the reference configurations.\\
Based on the idea of the Gaussian Overlap Approximation (GOA) \cite{ring}, 
we parameterize it as
\begin{equation}
N(q,q')=\exp(-\lambda(q-q')^2). 
\label{lambda}
\end{equation}
In the main calculations below, we take the value $\lambda=1.0$ for the overlap between 
neighboring $Q$-blocks. This sets the numerical scale for $q$
as a distance measure along the fission path. We also consider the model
in which the configurations are all orthogonal.

The Hamiltonian kernel $H(qj,q'j')$ can be calculated in a similar
manner by assuming that the orbital wave functions are nearly the same in
the two reference configurations.  To take into account the imperfect
overlap of the references states, we multiply the bare matrix elements
by the suppression factor $N(q,q')$
to the matrix elements.  
In addition, one has to take into account the
diabatic interaction between those configurations which are connected
diabatically. 
A simple formula for the diabatic interaction has been derived in 
Ref. \cite{hagino2022} based on a self-consistent separable interaction. 
Based on the GOA, the formula reads, 
\begin{equation}
\frac{\langle qj|v_{db}|q'j\rangle}{\langle
q|q'\rangle}=\frac{E(qj)+E(q'j)}{2}-h_2(q-q')^2,
\label{diabatic}
\end{equation}
where $E(qj)=k_j d + V(q)$ is the energy of the configuration $(qj)$. 
In the previous work \cite{BH2023}, the value of $h_2$ was estimated to be about 
1.5 MeV with the Gogny HFB calculations for $^{236}$U. 
A typical value of single-particle spacing $d$ is around 0.5 MeV when it is corrected for the effective mass 
(see Table III in Appendix A). Combining these together, we estimate $h_2=3d$ in this model.

The first term on the hand side of this equation
insures that the Green's function (\ref{green}) transforms properly
under a shift in energy scale $ E' = E - \epsilon$, that is 
$G'(E') = G(E)$. 

\subsection{Width matrices}

The matrices $\Gamma_a$ ($a$=$n$, cap, and fis) in Eq. (\ref{green}) can be in principle  
derived with the generalized Fermi Golden Rule\cite{bertsch2019}
\begin{equation}
(\Gamma_a)_{kk'}=2\pi\sum_{l\in a}\langle k|v|l\rangle \langle k'|v|l\rangle\delta(E_l-E)
\end{equation}
where $l$ labels states in the decay channel $a$.
Due to the non-orthogonality of the configurations, the matrix $\Gamma_a$ is
in general non-diagonal.  In this work, we take a separable approximation
and parameterize it as \footnote{In Ref.  \cite{CI2022}, we used $N$ instead
of $N^{1/2}$ in the decay matrices.  We consider that $N^{1/2}$ is a more
physical choice because of the connection to orthogonal bases as we discuss
in Appendix B.}
\begin{equation} 
(\Gamma_a)_{kk'}=\gamma_a \sum_{l \in a}(N^{1/2})_{k,l}(N^{1/2})_{k',l},
\label{gamma}
\end{equation} 
where $(N^{1/2})_{k,l}$ is the square root of the norm kernel and $\gamma_a$
is the mean decay width.  Here, the indices $k$ and $l$ label both the
deformation $q$ and the excitation $j$. See Appendix \ref{AppB} for a
derivation of Eq. (\ref{gamma}).

\section{RESULTS}

Let us now numerically evaluate the transmission coefficients and discuss
the dynamics of induced fission.  To this end, we consider a chain of
three $Q$-blocks, $q=q_{-1}, q_0$, and $q_1$, with the same spacing $\Delta
q$, that is, $q_{\pm 1}=q_0\pm\Delta q$.  We set  them to $q=-1, 0$, and 1 for
convenience.  Thus the overlaps between adjacent $Q$-blocks is 
$N(q,q\pm 1) = e^{-1}$ by Eq. (\ref{lambda}) with the chosen value of
$\lambda$. For the barrier, we set $V(q=\pm 1)=0$ and $V(q=0)=4d$, giving
a barrier height 
$B_h=4d$.  In each $Q$-block, the
energy cutoff for the many-body configurations is set to be $E_{\rm
cut}=V(q)+5.5d$.  The neutron absorption and the gamma decay occur prior to
the fission barrier, so the  incident and the capture channels couple to
the internal states by Eq. (\ref{gamma}) at $q=-1$. 
Likewise, the fission channel is coupled at $q=1$. All the states at these
end points are coupled to individual decay channels. 
Since the relation $\Gamma_n < \Gamma_{\rm cap} <
\Gamma_{\rm fis}$ is known empirically in the actinide region
\cite{bertsch2017}, we set $\gamma_n=0.001d, \gamma_{\rm cap}=0.01d$, and
$\gamma_{\rm fis}=0.1d$ in the following calculations.  As we will show in
Sec.  IIIC below, the transmission dynamics is not sensitive to the value of
$\gamma_{\rm fis}$.
					
\subsection{Orthogonal basis}

We first consider the case where 
all configurations are orthogonal so that the norm kernel reads 
\begin{equation}    
\langle q\, j |N| q'\, j'\rangle = \delta_{q,q'}\delta_{j,j'}. 
\end{equation}
In this case, the suppression factor in the off-diagonal couplings 
are disregarded, that is, the off-diagonal couplings are fully taken into account without the 
suppression factor.
This is a useful limit to study the role of the pairing interaction, since the
diabatic interaction does not contribute.

It is a well-known fact that the pairing correlation modifies drastically
the dynamics of spontaneous fission, particularly through a reduction of the
collective mass \cite{giuliani2014,sadhukhan2014,guzman2018}.  Another
important aspect of the pairing correlation is that it is responsible for a
hopping of Cooper pairs from one configuration to the neighboring one
\cite{barranco1990}.  On the other hand, the role of pairing correlation in
induced fission has not yet been understood well, partly because the pairing
correlation is considered to be effective only in the vicinity of the ground
state.  However, odd-even staggerings have been observed in fission
fragments in low-energy induced fission, which suggests that the pairing
correlation cannot be completely ignored.

\begin{figure}[t]
\includegraphics[width=8.6cm]{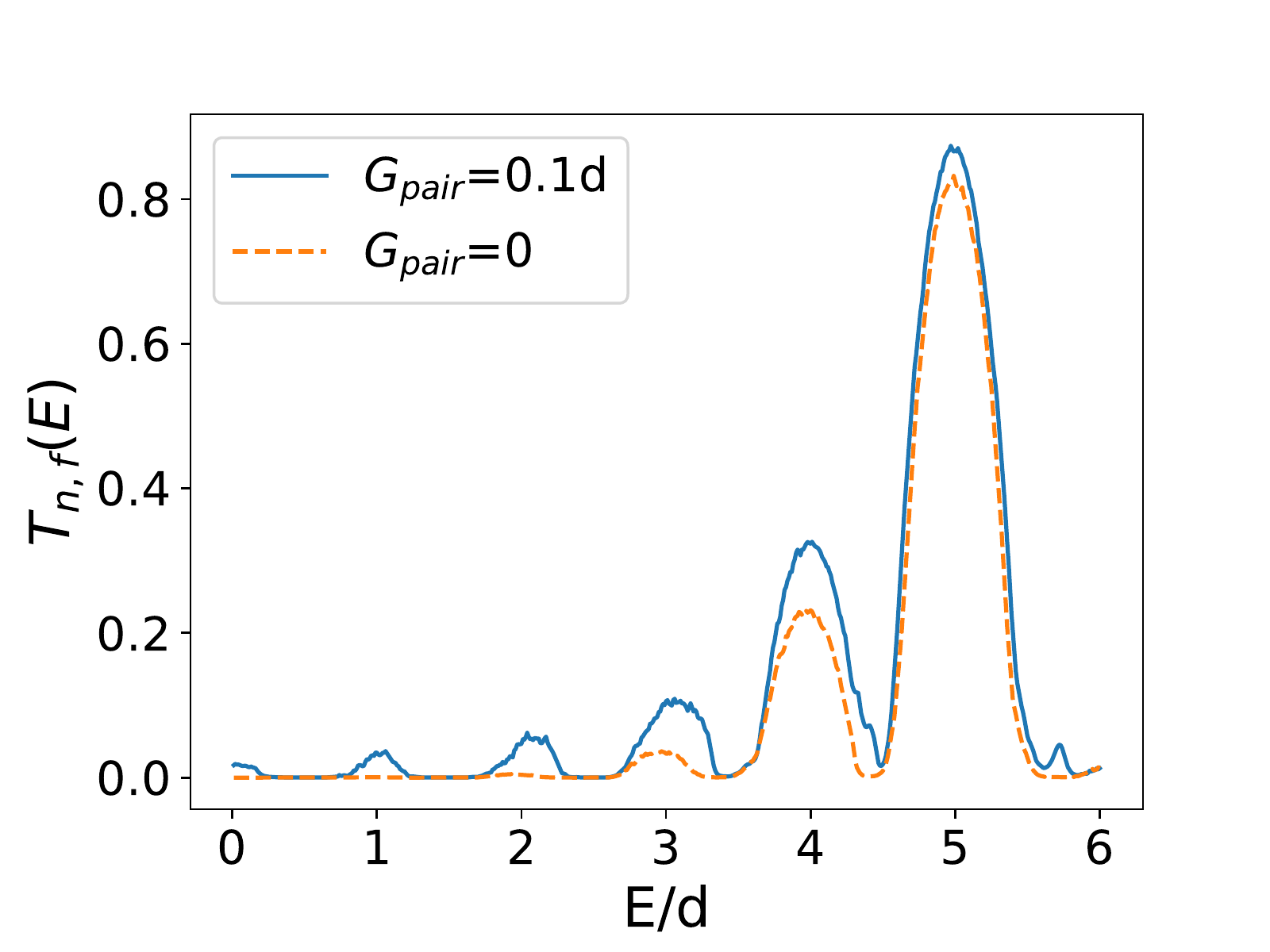}
\caption{The transmission coefficients from the incident channel to the fission channel
as a function of the excitation energy, $E$, in the model with
an orthogonal configuration space. The solid and the dashed lines are obtained with 
$G_{\rm pair}=0$ and $0.1d$, respectively, for the strength of the pairing interaction. 
The strength of the neutron-proton interaction and the barrier height are set to be 
$v_{np}=0.03d$ and $B_h=4d$, respectively. 
}
\end{figure}

Fig. 2 shows the transmission coefficients for the fission channel,
calculated  with 
two different values of $G_{\rm pair}$. The strength of the neutron-proton 
random interaction is set to be $v_{np}=0.03d$. 
One can see that 
the pairing correlation enhances the transmission probabilities far below the 
barrier, while its effect is not important at the barrier top and above.
This is to be expected, since the number of configurations with high 
seniority numbers increases 
as the excitation energy increases and the pairing correlation becomes weaker. 

\begin{table}[hbtp]

 \caption{The averaged transmission coefficient for a fission process, $\langle
 T_{n,{\rm fis}}\rangle$, for several sets of the interaction parameters and
 assuming that the configurations are orthogonal.   The barrier height
 and the incident energy are both set to be $4d$.  }
  \centering
  \begin{tabular}{c|c|ccc}
\hline
  \hline
    &        &     & $G_{\rm pair}$ &    \\
            \cline{3-5} 
  Model&$v_{np}$&0 &$0.1d$&$0.2d$\\
  \hline
  I&0&0&0.0441&0.0589 \\
  II&0.03d&0.107&0.161&0.173\\
  III&0.06d&0.318&0.331&0.331\\
  \hline
  \hline
  \end{tabular}
\end{table}

To study systematically the role of pairing in induced fission, 
we introduce an energy-averaged transmission coefficient.  It is  defined as 
\begin{equation}
\langle T_{n,{\rm fis}}(E)\rangle =\frac{1}{\Delta E}\int_{E-\Delta E/2}^{E+\Delta E/2} dE' \,T_{n,{\rm fis}}(E').
\label{average}
\end{equation}
Table I summarizes the energy averaged transmission coefficient at $E=B_h=4d$ for several sets of 
$(v_{np},G_{\rm pair})$. 
The energy window for the energy average is set to be $\Delta E=d$. 
Without the neutron-proton interaction, that is, $v_{np}=0$, 
the fission probability increases as the pairing strength increases. 
Note especially that the transmission coefficient $\langle T_{n,{\rm fis}}(E)\rangle$ 
is zero when there is no interaction at all. 
As the value of $v_{np}$ increases, the dependence of 
$\langle T_{n,{\rm fis}}(E)\rangle$ on $G_{\rm pair}$ becomes milder. 
For $v_{np}=0.06d$, the transmission coefficient is almost insensitive to the value of $G_{\rm pair}$. 
This suggests that induced fission is more sensitive to the neutron-proton random interaction, 
as compared to the coherent pairing interaction.  

\subsection{Non-orthogonal basis}

Let us now examine the dependence on the interactions for a model having a
non-orthogonal basis.  
In this case, 
diabatically connected configurations have a off-diagonal coupling due to the one-body terms in the Hamiltonian, 
in addition to the couplings due to the two-body residual interactions. 
This corresponds to the diabatic interaction (\ref{diabatic}) parameterized with a quantity $h_2$. 
To avoid an artifact due to the degeneracy of the single-particle energies, 
we introduce an offset energy to the $q=1$ block, taking 
$\vec V(q)/d = (0,4,0.5)$ in Eq. (\ref{hamiltonian}).
We mention that this problem appears much more prominently with the non-orthogonal basis 
as compared to calculations with the orthogonal basis, which could not be cured merely by introducing 
random numbers to the Hamiltonian kernel.

Figure 3 shows the transmission probability 
for fission  with two different values of $v_{np}$. 
In these calculations, the pairing interaction is switched off by setting $G_{\rm pair}=0$, while 
the parameter $h_2$ for the diabatic transitions is set to be 3$d$. 
From the figure, one notices that the peaks are lowered and broadened as the value of  
$v_{np}$ increases. 
This can be understood easily since the random interaction spreads 
the spectrum in each $Q$-block as is indicated in Fig. 1. 
The effect of $v_{np}$ is not only to broaden the peaks in the transmission 
coefficients but also to increase the energy averaged transmission coefficients, as 
will be discussed in Table II below. 

\begin{figure}
\includegraphics[width=8.6cm]{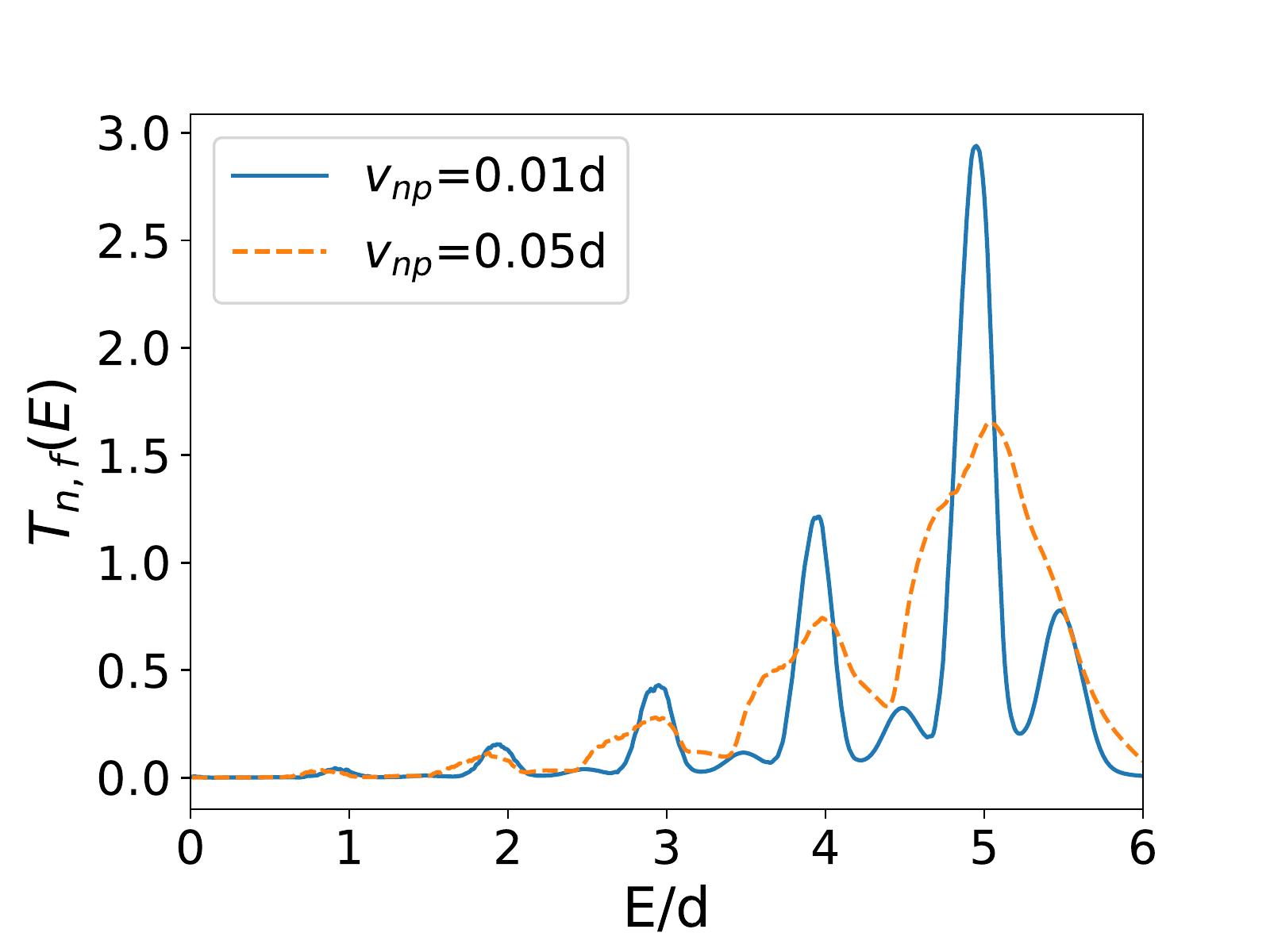}
\caption{
The transmission coefficient $T_{n,{\rm fis}}(E)$ with two different values of $v_{np}$.
The pairing interaction is set to zero, i.e., $G_{\rm pair}=0$. 
The other parameters are $\lambda=1.0,~h_2=3d$, and $B_h=4d$.
}
\end{figure}

Figure 4 shows an average fission-to-capture branching ratio $\alpha^{-1}$ as a function of the energy $E$. 
We define the average as 
\begin{equation}
\alpha^{-1}=\frac{\int dE' \,T_{n,{\rm fis}}(E')}{\int dE' \,T_{n,{\rm cap}}(E')},
\label{ratio}
\end{equation}
where the range of the integration is the same as that in Eq. (\ref{average}).
To simplify the discussion, we once again set the pairing interaction to be zero. 
{The solid and the dashed lines are obtained with $h_2=3d$ and $h_2=0$, respectively. 
The branching ratios increase with the excitation energy, and we confirmed that the energy dependence becomes stronger 
as the number of $Q$-blocks increases. 
This would be an expected behavior from a quantum barrier transmission. 
Furthermore, one sees that the diabatic interaction increases the branching ratios, that is consistent with the result in Ref. \cite{BH2023}.

\begin{figure}
\includegraphics[width=8.6cm]{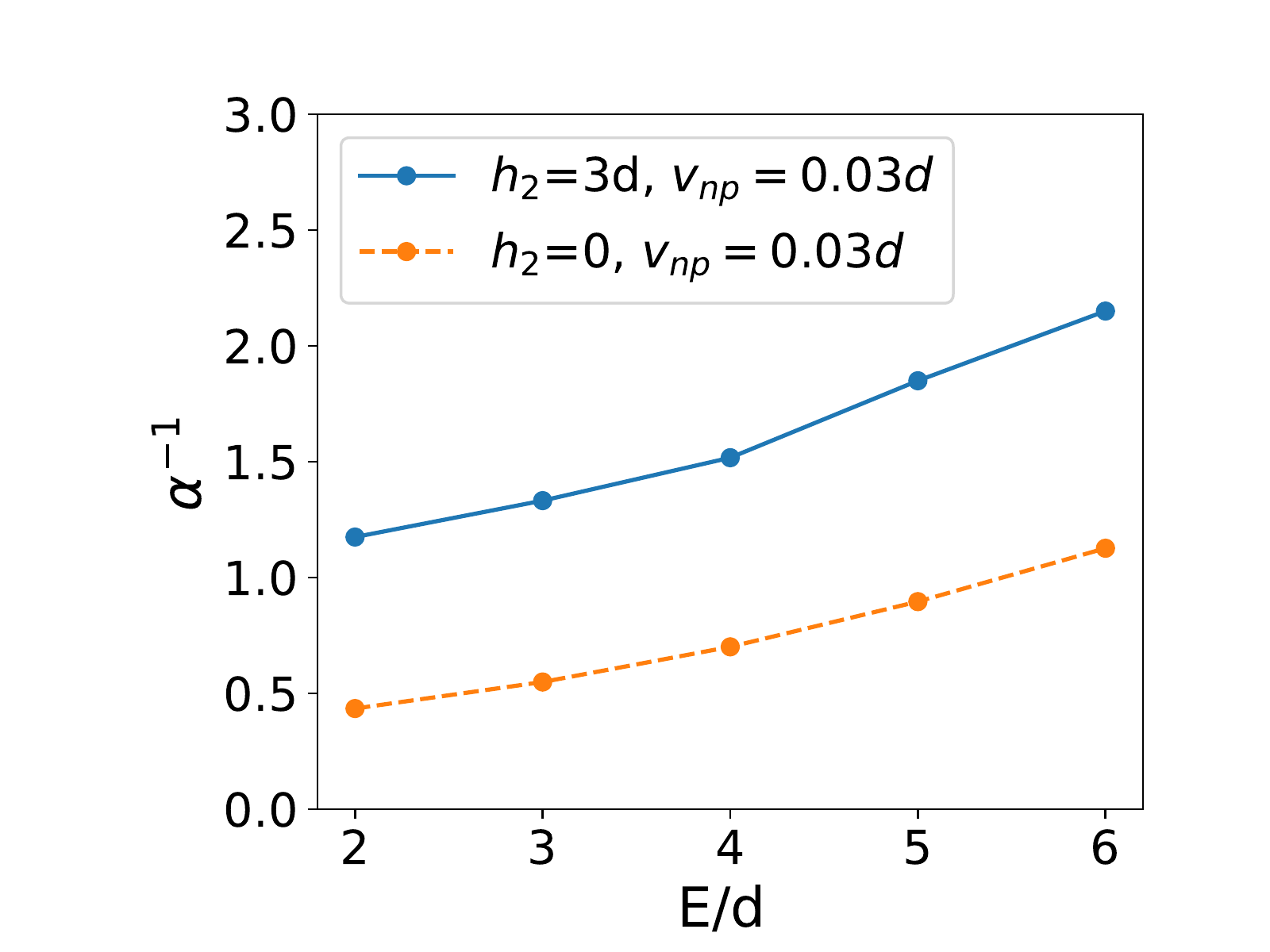}
\caption{
The average fission-to-capture branching ratios as a function of the energy with two different values of $h_2$.
The random neutron-proton interaction is taken into account with the strength of $v_{np}=0.03d$, while 
the pairing interaction is set to be zero. }

\end{figure}

\begin{table}[hbtp]
 \caption{
\label{tableII}
 The transmission coefficient for fission $\langle T_{n,{\rm fis}}\rangle$ and the
 branching ratio $\alpha^{-1}$ for several sets of interactions.
 The parameters shown for models I-V1 are the only ones that differ
 from the base model. The overlap parameter is 
 $\lambda=1.0$ and the averaged observables are calculated at a
 central energy $E = 4 d$.  Interaction strength parameters are in units of $d$.  For
 the model VI, the GCM Hamiltonian is constructed such that the
orthogonal physical Hamiltonian Eq. (\ref{3D-H0}) is diagonal.
}
  \centering
  \begin{tabular}{c|ccc|cc}
  \hline
   \hline
  Model&$v_{np}$&$G_{\rm pair}$&$h_2$&$\langle T_{n,{\rm fis}}\rangle$&$\alpha^{-1}$\\
  \hline
  base&0.03&0.3&3&0.417& 1.62\\  
  I&&0.0&&0.413& 1.49\\
   II&0.0& & &0.372&1.14 \\ 
  III&0.05& & &0.429&2.27 \\
  IV& & &0.0&0.294& 0.739\\
   V&0.0&0.0&0.0&0.172& 0.291\\ 
   VI&0.0&0.0&0.0&0.000& 0.000\\ 
  \hline
    \hline
  \end{tabular}
\end{table}

Table \ref{tableII} summarizes the transmission coefficients and the branching ratios
for several parameter sets.  The results of the models I, II, III indicate
that both the neutron-proton interaction and the pairing interaction enhance
the transmission coefficients as well as the branching ratios.  They also
indicate that the transmission coefficients are more sensitive to the
neutron-proton interaction than to the pairing interaction.  This is
consistent with the results of the orthogonal basis shown in Table I, even
though the degree of enhancement is smaller than in Table I due to the
overlap  factor $N(qj,q'j')$ in the off-diagonal matrix elements.  In
the model IV, the value of $h_2$ is set to be zero.  The result indicates
that the transmission coefficient and the branching ratio significantly
decreases without the diabatic transitions, as has been already observed in
Ref.  \cite{BH2023}.  See also Fig.  4 for the sensitivity of the branching
ratios to the value of $h_2$.  Finally, in the model V, all the interaction
strengths, $v_{np},~G_{\rm pair}$, and $h_2$, are set to be zero.  Even in this case,
the transmission coefficient is not zero, because the corresponding 
Hamiltonian in the orthogonal physical basis is not diagonal in this case. 
As we show in Appendix B, one can actually construct the GCM Hamiltonian 
which is diagonal with 
the orthogonal basis. 
With such a GCM Hamiltonian, we have confirmed that the transmission coefficient becomes zero within the
numerical error (see the model VI in the table).

\subsection{Validity of the transition state hypothesis}

In the Bohr-Wheeler theory for induced fission \cite{bohr1939}, the decay width is
calculated
as a sum of transmission coefficients $T_i$ across the barrier via transition states $i$,
\begin{equation}    
\Gamma_{\rm BW}=\frac{1}{2\pi\rho}\sum_iT_i,
\end{equation}
where $\rho$ is the level density of a compound nucleus.
The formula indicates that 
the transition states entirely determine the decay rate, and that 
the details of the dynamics after acrossing the barrier are unimportant. 
The branching ratio in the Bohr-Wheeler theory would be expressed as 
\begin{equation}
\alpha^{-1}(E)=\frac{1}{2\pi\rho(E)\Gamma_{\rm cap}}\sum_iT_i. 
\end{equation}

\begin{figure}
\includegraphics[width=8.6cm]{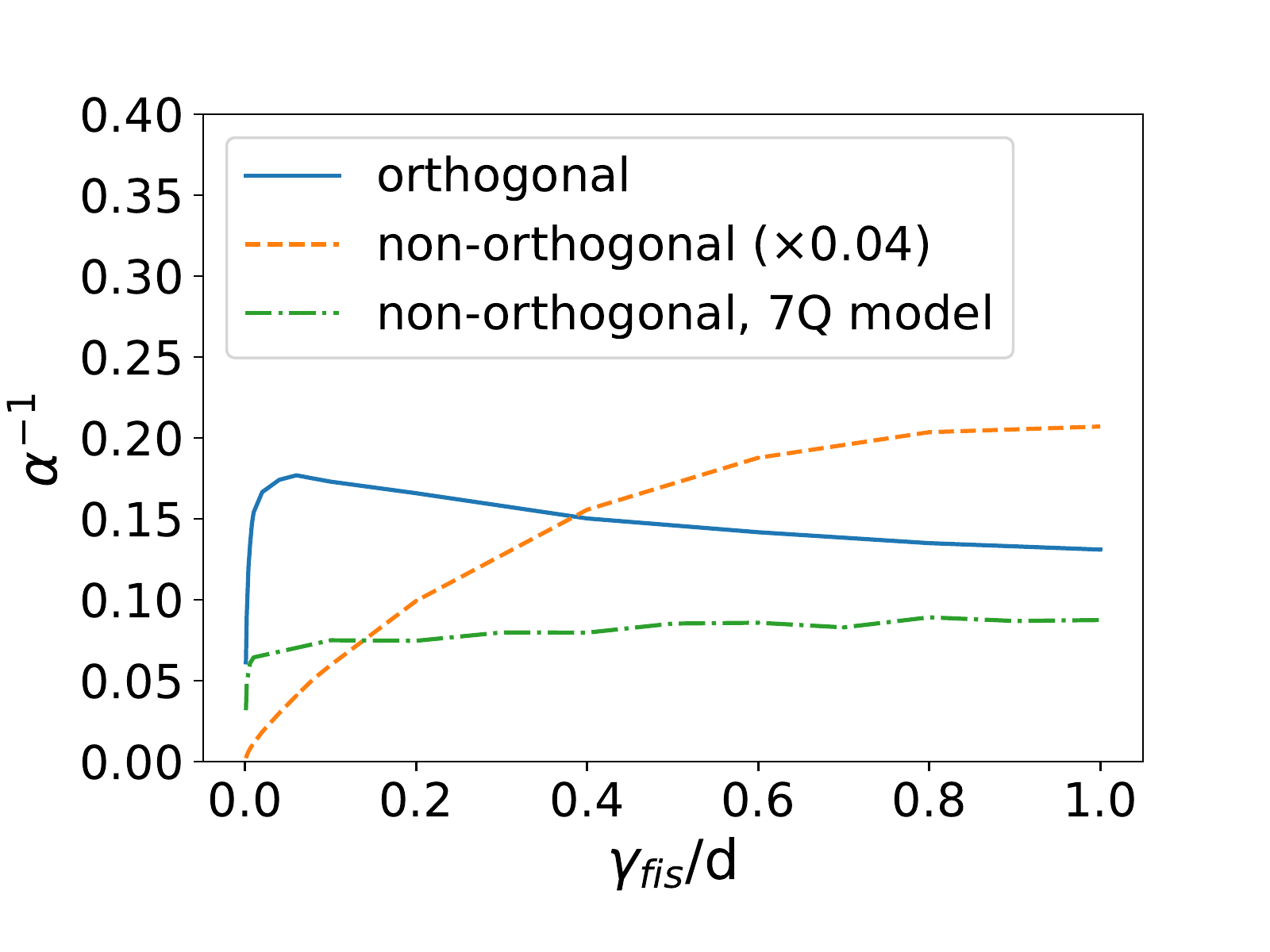}
\caption{
\label{alph_vs_gam}
The branching ratios at $E=4d$ as a function of $\gamma_{\rm fis}$.  The
interaction strengths are $(v_{pn},G_{\rm pair},h_2) = (0.03,0.3,3)d$.
A parabolic fission
barrier is employed with the barrier height of 4$d$.  The solid and the
dashed lines show the results of the 3$Q$ model, while the dot-dashed line
the results of the 7$Q$ model. While the non-orthogonality of the
configurations is neglected in the solid line, it is taken into account in
the other lines with $\lambda=1.0$. 
For the sake of presentation, the branching ratios are multiplied by factors of 0.2 and 0.04 for the solid and the dashed lines, respectively.
}
\end{figure}
The solid and the dashed lines in Fig.  \ref{alph_vs_gam} show the branching ratios at
$E=4d$ as a function of $\gamma_{\rm fis}$ for a model with the
pairing interaction switched off. 
For the calculations with the orthogonal basis shown by
the solid line, the branching ratio is almost independent of the fission
decay width $\gamma_{\rm fis}$, in agreement with the insensitivity property
of the Bohr-Wheeler theory.  On the other hand with the non-orthogonal basis, the
branching ratio increases gradually as a function of $\gamma_{\rm fis}$,
even though the insensitivity property may be realized at large values of
$\gamma_{\rm fis}$.  To check the dependence on the number of $Q$-blocks, we
repeat the calculations with 7 $Q$-blocks, parameterizing $V$ as
$V(q)/d=4-4q^2/9$ ranging from $q=-3$ to $q=3$ with
$\Delta q=1$.  
  In this case, the branching
ratio changes by less than a factor of two while the fission decay
varies by an order of magnitude.
All of these results indicate that the hypothesis used in the
Bohr-Wheeler theory is easily realized in the present microscopic theory. 
See also Ref.  \cite{CI2021B} for a similar study with random matrices.

\section{Summary}

In this article, we have applied the CI methodology to a schematic
model for neutron-induced fission.  The model Hamiltonian 
contains the pairing interaction,
the diabatic interaction, and a schematic off-diagonal neutron-proton
interaction.  The
model appears to be sufficiently detailed to examine the sensitivity of the
fission transmission probabilities to the different types of interaction, as well
as the validity of transition state theory in a microscopic framework.  We
have shown that the transmission coefficients are mainly sensitive to the
neutron-proton interaction, while the sensitivity to the pairing interaction
is much milder.  The diabatic transitions were also found to play a role. 
Depending on the interaction and the deformation-dependent configuration
space, one achieves conditions in which branching ratios depend largely on
barrier-top dynamics and are insensitive to properties closer to the
scission point.  The insensitive property is one of the main assumptions in
the well-known Bohr-Wheeler formula for induced fission, but up to now it
had no microscopic justification.

The results in this paper indicate that the neutron-proton interaction is an
important part of a microscopic theory for induced
fission.  To include it in realistic calculations based on the density
functional theory will require a large model space, however.  See Table III
in Ref. \cite{BH2023} for some estimates of the dimensional requirements. 
Moreover, single-particle energies are in general not degenerate in contrast to 
the schematic model employed in this paper.  This might require a different energy cutoff, further 
enlarging the model space.
To carry out
such large scale calculations for induced fission, one will have to either
validate an efficient truncation scheme or develop an efficient numerical
method to invert matrices with large dimensions.  We leave this for a future
work.

\begin{acknowledgments}
We are grateful to G.F. Bertsch for collaboration on the early stage of this work. 
This work was supported in part by JSPS KAKENHI
Grants No. JP19K03861 and No. JP21H00120.
This work was supported by JST, the establishment of university fellowships towards
the creation of science technology innovation, Grant Number JPMJFS2123. 
The numerical calculations were performed
with the computer facility
at the Yukawa Institute for Theoretical
Physics, Kyoto University.
\end{acknowledgments}

\appendix

\section{Estimation of physical parameters}

\subsection{Orbital energy spacing}
The single-particle level spacing $d$ in the uniform model
sets the energy scale for the model and does not
play any explicit role in the model.  However, it is required to
determine other energy parameters which are expressed
in units of $d$.  Several estimates of $d$ for \uf~ 
are given in Table III.  The first is based on orbital energies
in a deformed Woods-Saxon potential with the parameters given 
in Ref. \cite{bm1}; see Table IV for the
calculated orbital energies.
\begin{table}[htb] 
\begin{center} 
\begin{tabular}{|c|c|} 
\hline
$d$ (MeV) &  Source \\
\hline
0.45  &  Woods-Saxon well \\
0.51 &  FRLDM  \cite{mo95} \\
0.33 &  FGM  \cite{ko08} \\
\hline
\end{tabular} 
\caption{Estimated orbital level spacing in \uf.  The first
two are from potential models and the last extracted from
the Fermi gas formula and measured level densities.
}
\label{orbital-d} 
\end{center} 
\end{table} 
\begin{table}[htb] 
\begin{center} 
\begin{tabular}{|ccc|ccc|} 
\hline 
\multicolumn{3}{|c|}{protons} & \multicolumn{3}{c|}{neutrons}\\
\hline
$2K$ & $\pi$ & $\varepsilon_{K\pi}$ (MeV) & $2K$ & $\pi$ & $\varepsilon_{K\pi}$ (MeV) \\
\hline
3 & $-$1 & $-$3.39 & 5 & $-$1  & $-$4.15 \\
5 & $-$1 & $-$3.80& 1 & $-$1 & $-$4.25 \\
5 &  1& $-$4.93 & 7 & $-$1 & $-$4.40 \\
\multicolumn{3}{|c|}{- - - - - -} & \multicolumn{3}{c|}{- - - - - -}\\
1 & 1& $-$5.43 & 1 & 1 & $-$5.07  \\
9 & $-$1& $-$5.53 & 5 & 1 & $-$5.75 \\
3 &  1 & $-$5.74 & 5 & $-$1 & $-$5.82 \\
\hline
\hline\end{tabular} 
\caption{Characteristics of single-particle orbitals in a deformed
Woods-Saxon potential corresponding to \uf~ at deformation
$(\beta_2,\beta_4) = (0.274,0.168)$. Dashed line indicates the Fermi level.
}
\label{orbital-e} 
\end{center} 
\end{table} 
In
more realistic theory, the momentum dependence of the potential
tends to increase the spacing, but the coupling to many-particle
degrees of freedom decreases the spacing of the quasiparticle 
poles.  The combined effect seems to  somewhat decrease
the spacing\footnote{We note that an energy density functional
fitted to fission data\cite{ko12} obtained an  effective mass in the single-particle
Hamiltonian  very close to  1.}\\  

\subsection{Level density} 
It is important to know the composition 
of the levels in the
compound nucleus to construct microscopic models that involve
those levels.  For a concrete example, consider the levels at
the neutron threshold energy $S_n = 6.5$ MeV in $^{236}$U.
The predominating configurations at this energy should be
$k$ subblocks at $k\approx S_n/d$ in the independent quasiparticle
approximation.  Another approach that is less sensitive to the
residual interaction is to estimate the total number of states
below $S_n$ and compare it to the number obtained by summing
the $N_k$ degeneracies in the $Q$-block spectrum.  In the
\uf~ example, the combined level spacing of $J^\pi = 3^-$ and
$4^-$ is about 0.45 eV at $S_n$\cite{ca09}.  At that excitation energy
the level density is the same for even and odd parities, and it
varies with angular momentum as $2J+1$.  The inferred 
level spacing of $J^\pi = 0^+$ levels is thus about 7 eV. The
accumulative number of levels can be approximated  by
$N = \rho T$ where $T$ is the nuclear temperature, defined as
$T = d \log(\rho(E))/d E$.  A typical estimate for our example
is $T= 0.65$ MeV,  giving $N \approx 1.0\times10^8$. 
To estimate the level density in the present model, we start
with the set of quasiparticle configurations including both
parities and all $K$ values.  The resulting $k$-blocks have
multiplicities that are well fit by the formula 
\be
N_k \approx \exp(-3.23 + 4.414 k^{1/2}).
\ee 
Projection on good parity decreases this by a factor of two.  The
projection on angular momentum $J=0$ is more subtle.
The $J=0$
states are constructed by projection from $K=0$ configurations; other
configurations do not contribute.  However, there may be two distinct
configurations that project to the same $J=0$ state.  This gives
another factor of nearly two reduction in the multiplicity.  The
remaining task is to estimate the fraction of $K=0$ configurations
in the unprojected quasiparticle space.  The distribution of 
$K$ values is approximately Gaussian with a variance given by
\be
\langle K^2\rangle = \langle n_{\rm qp}\rangle \langle K^2\rangle_{\rm sp}
\ee
where $\langle n_{\rm qp}\rangle\approx 8$  is the average number of quasiparticles
in the $k$ block and $\langle K^2\rangle_{\rm sp} \approx 6$ is an
average over the orbital $K$'s near the Fermi level.  Including
these projection factors, the integrated number of levels up
to $S_n$ is achieved by including all $k$-subblocks up to $k=17$
in the entry $Q$-block. \\

\subsection{Neutron-proton interaction}
\label{A3}
To set the scale for our neutron-proton interaction parameter
$v_{\rm np}$ we compare it with phenomenological contact interactions
that have been used to model nuclear spectra.  The matrix 
element of the neutron-proton 
interaction is
\be
\langle n_1 p_1 | v | n_2 p_2\rangle = -v_0 I
\ee
where
\be
I = \int d^3 r   \phi_{n_1}^*(\vec r)\phi_{p_1}^*(\vec r)
\phi_{n_2}(\vec r) \phi_{p_2}(\vec r).
\label{eq-I}
\ee
The parameter $v_0$ is the strength of the interaction, typically
expressed in units of MeV fm$^3$.  
Some values
of $v_0$ from the literature are tabulated in Table \ref{v0}.  
\begin{table}[htb] 
\begin{tabular}{|c|c|c|} 
\hline 
Basis of estimate  &  $v_0$ (MeV fm$^3$) &  Citation \\
\hline 
$G$-matrix    & 530  &  \cite{BBB92} \\
$sd$-shell spectra & 490 & \cite{ba88}\\        
$\beta$-decay & 395,320 & \cite{yo13}\\
\hline 
\end{tabular} 
\caption{Estimates of neutron-proton interaction strength.
\label{v0} 
}
\end{table} 
We shall adopt the value $v_0 = 500 $ MeV fm$^3$ 
to estimate the value of $v_{np}$.

If the wave functions
of the eigenstates approach the compound nucleus limit,
the only characteristic we need to know is its mean-square
average among the active orbitals.  We have used the Woods-Saxon model to 
calculate the 
integral Eq. (\ref{eq-I}) for all the fully off-diagonal matrices
of the orbitals within 2 MeV of the Fermi energy. 
Fig. \ref{v_np-hist} shows a histogram of their distribution 
\footnote{If the 
orbitals are restricted only to those in TABLE IV, the histogram is more 
structured.}.
\begin{figure}
\includegraphics[width=1.0\columnwidth]{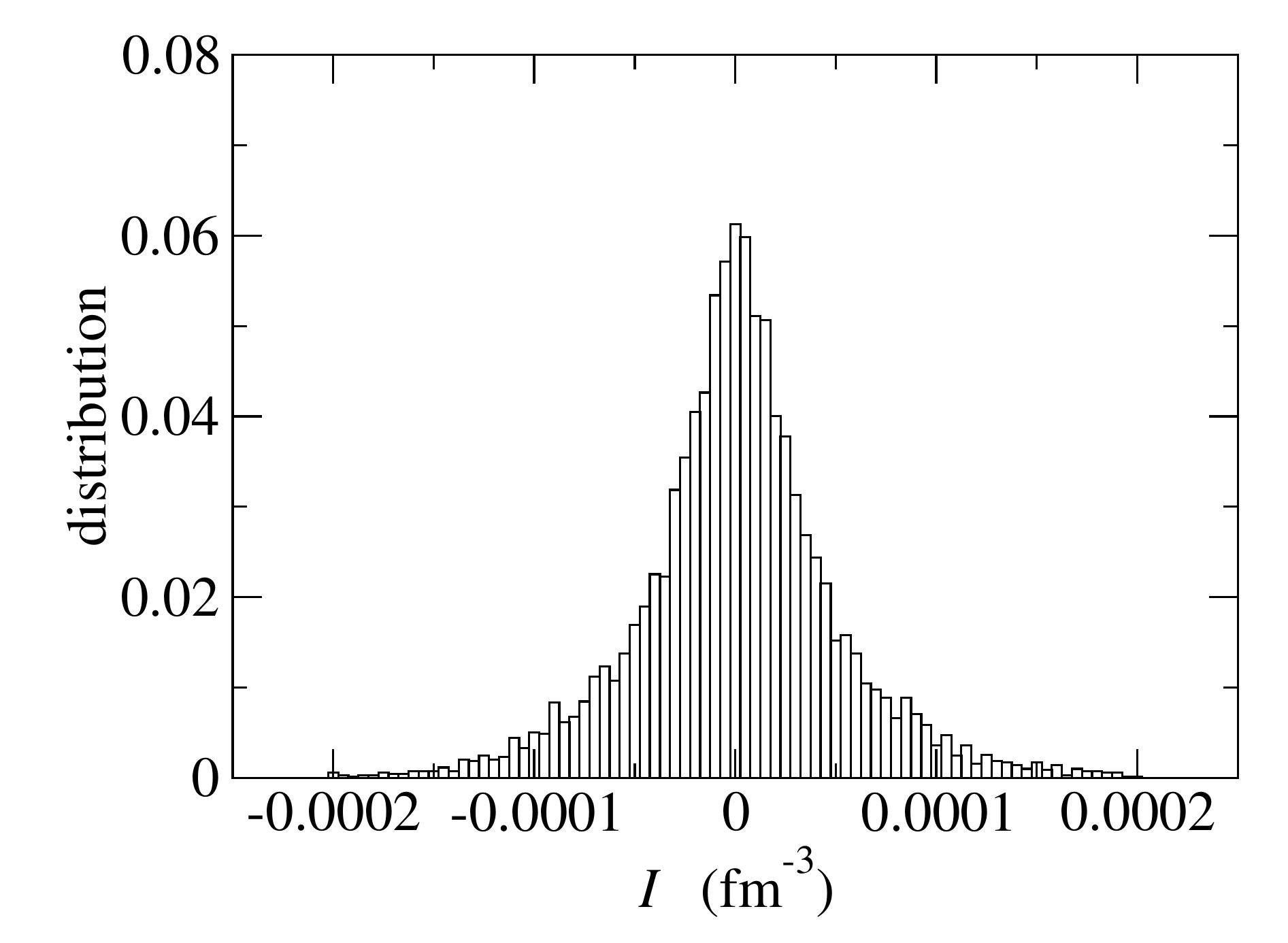}
\caption{Integrals $I$  in Eq. (\ref{eq-I}) of orbitals near
the Fermi energy.  
\label{v_np-hist}
}
\end{figure}
The variance of the distribution is 
$\langle I^2\rangle^{1/2} = 5.22\times 10^{-5}$ fm$^{-3}$.  
Combining this with our estimate of $v_0$ we find 
$(\overline{\langle n_1 p_1 |v |n_2 p_2\rangle^2})^{1/2} = 0.025$ MeV.
This implies $v_{\rm np} \sim 0.05 d$ with our estimated single-particle
level density.\\

\section{Reaction theory in a non-orthogonal
basis}
\label{AppB}
The space of configurations used in this work is not
orthogonal.  This causes some conceptual issues, but 
it does not cause a significant computational 
burden in CI-based reaction theory.  
The theory is based on calculating the resolvent of $H$;
in an orthogonal basis it is given by 
\be
\label{G}
G = (H - E \unit)^{-1}
\ee  
where $\unit$ is the unit matrix, $H$ is the Hamiltonian, and
$E$ is the energy of the reaction. 
Non-orthogonal bases 
also arise in the theory of spontaneous decays \cite{ha20}, and
in electron transport theory when wave functions are built from atomic orbitals.
See for example Refs. \cite{re11,ho07,ho06,pe04,na99} for the
formulation of the resolvent as
commonly used  in chemistry and condensed matter physics. 

In a non-orthogonal basis the time-dependent Schr\"odinger equation
reads
\be
H \Psi = i \hbar  N\, \frac{d}{dt}\, \Psi
\ee
where $N$ is the overlap matrix between basis states
$N_{ij} = \langle i | j \rangle$.  The corresponding
resolvent is 
\be
G = (H - E N )^{-1}.
\ee
There is hardly any difference from Eq. (\ref{G}) from a 
computational point of view.  However, the couplings to reaction
channels should be treated with care.
\def\vgcm{\vec v_{\rm gcm}}
\def\vph{\vec v_{\rm phys}}

To understand the couplings, we define a certain
orthogonal basis which we call the physical basis.
We call the vector representing a wave
function in that basis 
 $\vph$ and in the 
non-orthogonal basis as $\vgcm$. In the 
GCM the dot products of basis elements satisfy 
\be
\vgcm(i)^* \cdot \vgcm(j) = N_{ij}
\label{S}
\ee
while those  in the physical basis satisfy
\be
\vph(i)^* \cdot \vph(j) = \delta_{ij}.
\ee
A physical basis consistent with Eq. (\ref{S}) can then 
be defined by setting
\be
\vph(i) = \sum_j N^{1/2}_{ij} \vgcm(j).
\ee
This definition is not unique since the dot products
are invariant under a unitary transformation of the
physical basis.  Indeed, an orthogonal 
basis is usually constructed in the GCM by diagonalizing $N$ and using
its eigenvectors as the basis.   However, those basis states 
are not well localized with respect to the
GCM coordinate.

The relationship between the Hamiltonians in the
physical and GCM bases can be expressed
\be
\tilde{H} = N^{-1/2} H N^{-1/2}
\label{Hph}
\ee
or 
\be
H = N^{1/2} \tilde{H}  N^{1/2}.
\ee
The physical resolvent is related to
the GCM resolvent by
\be
\label{Gphysical}
\tilde{G} = \left( N^{-1/2} H N^{-1/2} - E\unit\right)^{-1}
\ee
$$
  = N^{1/2} \left(H - E N\right)^{-1}N^{1/2}. 
$$
One see that the matrix inversion is the same as in Eq. (\ref{G}) 
except for the replacement  $ \unit \rightarrow N$.  However,
the matrix $N^{1/2}$ appears as pre- and post-factors.

In our applications of CI-based reaction theory we assume that each
channel is coupled to a single state (the \lq\lq doorway" state) in the
internal space.  Taking that state to be the  basis state $d$ in the physical
representation,  the decay coupling matrix $\Gamma$ has
elements \footnote {A somewhat similar formula was used in Ref. \cite[Eq.
16]{CI2022}.}
\begin{equation}
\Gamma(i,j)=N^{1/2}_{id}\, N^{1/2}_{jd}\tilde{\Gamma}
\label{Ggcm}
\end{equation}
where $\tilde{\Gamma}$ is the decay width of the physical state $d$
into the channel.
Note that with this construction the transmission coefficient in the
physical basis
\begin{equation}    
T_{a,b}=\mathrm{Tr}[\tilde{\Gamma}_a\tilde{G}(E)\tilde{\Gamma}_b\tilde{G}^{\dagger}(E)]
\end{equation}
is transformed to 
\begin{equation}    
T_{a,b}
=\mathrm{Tr}\left[\Gamma_a G(E)\Gamma_bG^\dagger(E)\right]
\end{equation}
in the GCM basis.

There is another reason for explicit construction of the physical basis.
The distinction between the GCM and  physical basis
must be taken into account 
in Sec. IIIB where we assessed the relevant
importance of different interaction types and we want to start with
a Hamiltonian $\tilde{H}^0$ for which the transmission probability
vanishes.  One cannot simply set the off-diagonal elements of 
$H$ to zero if the overlap matrix $N$ connects
the entrance and exit channels, even if the connection is indirect.
It is the physical Hamiltonian $\tilde{H}$ that must be diagonal.
In two dimensions the construction is obvious.  Given
the diagonal elements of $H(i,i) = E_i$, the Hamiltonian
that is diagonal in the physical basis is
\be
\label{H0}
H^0 = \left(\begin{matrix}
     E_1 & (E_1 + E_2) N_{12} /2 \cr
     (E_1 + E_2) N_{12} /2 & E_2
           \end{matrix}\right).
\ee
Eq. (\ref{H0}) can be viewed as a justification for the first
term in Eq. (\ref{diabatic}). 
The construction can be carried out in higher dimensions using only linear
algebra operations, but we have no simple formula for the off-diagonal
elements of $H^0$.  For the base Hamiltonian treated in Sect.
IIIB, $N$ is given by
\be
N = \left(\begin{matrix}
     1.0 & e^{-1} & e^{-4} \cr
     e^{-1} & 1.0 & e^{-1} \cr
     e^{-4} & e^{-1} & 1.0 
           \end{matrix}\right).
\ee
Keeping only $V(q)$ in $H$, the $H^0$ is numerically found to be 
\be
\label{3D-H0}
H^0  = \left(\begin{matrix}
     0.0000 & 0.7571 & 0.1537 \cr
     0.7571 & 4.000 & 0.8547 \cr
     0.1537 & 0.8547 & 0.5000 \cr
           \end{matrix}\right).
\ee


\begin{thebibliography}{99}

\bibitem{hahn1939}O. Hahn and  F. Strassmann, Naturwissenschaften {\bf 27}, 11 (1939).
\bibitem{va73} R. Vandenbosch and J.R. Huizenga, {\it Nuclear Fission},
(Academic Press, New York, 1973).
\bibitem{bohr1939}N. Bohr and J. A. Wheeler, Phys. Rev. {\bf 56}, 426 (1939).
\bibitem{hauser1952}W. Hauser and H. Feshbach, Phys. Rev. {\bf 87}, 366 (1952).
\bibitem{randrup2011} J. Randrup and P. M\"oller, Phys. Rev. Lett. {\bf 106}, 132503 (2011).
\bibitem{aritomo2014}Y. Aritomo, S. Chiba, and F. Ivanyuk, Phys. Rev. C{\bf 90}, 054609
(2014).
\bibitem{ishizuka2017}C. Ishizuka, M. D. Usang, F. A. Ivanyuk, J. A. Maruhn, K. Nishio, and S. Chiba, Phys. Rev. C {\bf 96}, 064616 (2017).
\bibitem{schmidt2018}K.H. Schmidt and B. Jurado, Rep. Prog. Phys. {\bf 81}, 106301 (2018).
\bibitem{bender2020}M. Bender et al., J. Phys. G: Nucl. Part. Phys. {\bf 47}, 113002 (2020).
\bibitem{CI2022}G.F. Bertsch and K. Hagino, Phys. Rev. C {\bf 105}, 034618 (2022).
\bibitem{BH2023}G.F. Bertsch and K. Hagino, arXiv: 2302.00572 (2023).
\bibitem{ring}P. Ring and P. Schuck, {\it The Nuclear Many-Body Problem} (Springer-Verlag, Berlin, 2000).
\bibitem{donau1989}F. D\"onau, J. Zhang and L. Riedinger, Nucl. Phys. {\bf A496}, 333 (1989).
\bibitem{CI2021A}G.F. Bertsch and K. Hagino, arXiv:2102.07084 (2022).

\bibitem{datta}S. Datta, {\it Electronic Transport in Mesoscopic Systems} (Cambridge University Press, Cambridge, 1995).
\bibitem{al20} Y. Alhassid, G.F. Bertsch, and P. Fanto, Ann. Phys. (N.Y.) {\bf 419},
168233 (2020).
\bibitem{al21} Y. Alhassid, G.F. Bertsch, and P. Fanto, Ann. Phys. (N.Y.) {\bf 424},
168381 (2021).
\bibitem{martinez-larraz2022}J. Mart\'inez-Larraz and T.R. Rodr\'iguez, 
Phys. Rev. C{\bf 106}, 054310 (2022).
\bibitem{superfluidity}D.M. Brink and R.A. Broglia,
{\it Nuclear Superfluidity: Pairing in Finite System} (Cambridge University Press, 2005).
\bibitem{CI2020}G.F. Bertsch, Phys. Rev. C {\bf 101}, 034617 (2020).
\bibitem{bush1992}B.W. Bush, G. F. Bertsch and B. A. Brown, Phys. Rev. C {\bf 45}, 1709 (1992).
\bibitem{hagino2022}K. Hagino and G. F. Bertsch, Phys. Rev. C {\bf 105}, 034323 (2022).
\bibitem{bertsch2019}G.F. Bertsch and L.M. Robledo, Phys. Rev. C {\bf 100}, 044606 (2019).
\bibitem{bertsch2017}G.F. Bertsch and T. Kawano, Phys. Rev. Lett. {\bf 119}, 222504 (2017).
\bibitem{giuliani2014}S. A. Giuliani, L. M. Robledo, and R. Rodriguez-Guzman, Phys. Rev. C {\bf 90}, 054311 (2014).
\bibitem{sadhukhan2014}J.Sadhukhan, J. Dobaczewski, W. Nazarewicz, J. A. Sheikh, and A. Baran, Phys. Rev. C {\bf 90}, 061304(R) (2014).
\bibitem{guzman2018}R. Rodriguez-Guzman and L. M. Robledo, Phys. Rev. C {\bf 98}, 034308 (2018).
\bibitem{barranco1990} F. Barranco, G. Bertsch, R. Broglia, and E. Vigezzi, Nucl. Phys. {\bf A512}, 253 (1990).
\bibitem{CI2021B}G.F. Bertsch and K. Hagino, J. Phys. Soc. Jpn. {\bf 90}, 114005 (2021).
\bibitem{bm1}A. Bohr and B.R. Mottelson, {\it Nuclear Structure} (W.A.
Benjamin, Reading, MA, 1969), Vol. I.
\bibitem{mo95}P. M\"oller, et al., Atomic Data and Nuclear Data Tables
{\bf 59} 185 (1995); private communication (P. M\"oller).
\bibitem{ko08}A.J. Koning, S. Hilaire, and S. Goriely, Nucl. Phys. A {\bf 810} 13 (2008).
\bibitem{ko12}M. Kortelainen, et al., Phys. Rev. C {\bf 85} 024304 (2012).
\bibitem{ca09}R. Capote, et al., Nucl. Data Sheets {\bf 110} 3107 (2009).
\bibitem{BBB92} B.W. Bush, G.F. Bertsch and B.A. Brown, Phys. Rev. {\bf C45}, 1709 (1992).
\bibitem{ba88}B.A. Brown, et al., Ann. Phys. {\bf 182} 191 (1988).
\bibitem{yo13}K. Yoshida, Prog. Theor. Exp. Phys. 113D02 (2013).
\bibitem{ha20} K. Hagino and G.F. Bertsch, Phys. Rev. C {\bf 102} 024316
\bibitem{re11} M.G. Reuter, T. Seideman, and M.A. Ratner, Phys.
Rev. B {\bf 83} 085412 (2011). 
\bibitem{ho07} O. Hod, et al., Phys. Rev. B {\bf 76} 233401 (2007).	
\bibitem{ho06} O. Hod, J. Peralta, and G.E. Scuseria, J. Chem. Phys.
{\bf 125} 114	704 (2006)
\bibitem{pe04} A. Pecchia and A. Di Carlo, Rep. Prog. Phys. {\bf 67}
1497 (2004).
\bibitem{na99} M.B. Nardelli, Phys. Rev. B {\bf 60} 7828 (1999).

\end{thebibliography}
\end{document}